\documentclass[useAMS,usenatbib,usegraphicx]{mn2e}  
\usepackage[bottom]{footmisc}
\usepackage{color,amsmath,enumitem}
\usepackage{threeparttable}
\usepackage{graphicx}
\DeclareGraphicsExtensions{.pdf,.eps,.jpg,.ps,.eps}
\usepackage{epsfig}
\usepackage{url}
\usepackage{ amssymb }

\begin{document}

\title[Bayesian Analysis of Two Populations in 30 Galactic Globular Clusters]{Bayesian Analysis of Two Stellar Populations in Galactic Globular Clusters III: Analysis of 30 Clusters}

\author[Wagner-Kaiser et al.]{R. Wagner-Kaiser$^{1}$, D. C. Stenning$^{2}$, A. Sarajedini$^{1}$, T. von Hippel$^{3}$,  D. A. van Dyk$^{4}$, 
\newauthor{E. Robinson$^{5}$, N. Stein$^{6}$, W. H. Jefferys$^{7, 8}$} \\
$^1$Bryant Space Center, University of Florida, Gainesville, FL \\
$^2$Sorbonne Universit\'{e}s, UPMC-CNRS, UMR 7095, Institut d'Astrophysique de Paris, F-75014 Paris, France \\
$^3$Center for Space and Atmospheric Research, Embry-Riddle Aeronautical University, Daytona Beach, FL \\
$^4$Imperial College London, London, UK \\
$^5$Argiope Technical Solutions, FL \\
$^6$The Wharton School, University of Pennsylvania, Philadelphia, PA \\
$^7$University of Texas, Austin, TX, USA \\
$^8$University of Vermont, Burlington, VT, USA \\ }
\date{}

\pagerange{\pageref{firstpage}--\pageref{lastpage}} \pubyear{2014}

\maketitle

\label{firstpage}


\begin{abstract}
We use Cycle 21 Hubble Space Telescope (HST) observations and HST archival ACS Treasury observations of 30 Galactic Globular Clusters to characterize two distinct stellar populations. A sophisticated Bayesian technique is employed to simultaneously sample the joint posterior distribution of age, distance, and extinction for each cluster, as well as unique helium values for two populations within each cluster and the relative proportion of those populations. We find the helium differences among the two populations in the clusters fall in the range of $\sim$0.04 to 0.11. Because adequate models varying in CNO are not presently available, we view these spreads as upper limits and present them with statistical rather than observational uncertainties. Evidence supports previous studies suggesting an increase in helium content concurrent with increasing mass of the cluster and also find that the proportion of the first population of stars increases with mass as well. Our results are examined in the context of proposed globular cluster formation scenarios. Additionally, we leverage our Bayesian technique to shed light on inconsistencies between the theoretical models and the observed data.

\noindent{\it Keywords}:  (Galaxy:) globular clusters: general, (stars:) Hertzsprung-Russell and colour-magnitude diagrams, Galaxy: formation

\end{abstract}

\vspace{2pc}


\section{Introduction}\label{Intro}

The historical understanding of globular clusters as simple stellar populations has changed quickly and dramatically in the past decade. Overwhelming evidence has amassed that globular clusters are not simply a singular homogenous assembly of stars; rather, the distinct stellar groupings of chemical characteristics within the clusters are interpreted as multiple discrete populations (e.g.: \citealt{Bedin:2004}, \citealt{Gratton:2004}, \citealt{Carretta:2006}, \citealt{Villanova:2007}, \citealt{Piotto:2007}, \citealt{Piotto:2009}, \citealt{Milone:2009}, \citealt{Milone:2012a}). In order to better understand the history and evolution of the Milky Way, it is vital to understand these clusters and their properties.

New data from \cite{Piotto:2015} shows the vast majority of Galactic globular clusters host multiple populations. The immense influx of new data has spurred many new studies, both observational and theoretical, on the characteristics of the clusters (\citealt{Bedin:2004}, \citealt{Gratton:2004}, \citealt{Carretta:2006}, \citealt{Villanova:2007}, \citealt{Piotto:2007}, \citealt{Piotto:2009}, \citealt{Milone:2009}, \citealt{Milone:2012a}, \citealt{Piotto:2015}, \citealt{Nardiello:2015}, among others). Differences in helium abundances and the light elements carbon, nitrogen, and oxygen (CNO) produce visible effects seen in ultraviolet color-magnitude diagrams. However, the source of these varying abundances is still unclear, though several possible scenarios have been suggested. Among the proposed mechanisms are asymptotic giant branch (AGB) stars and fast-rotating massive stars (FRMS), where later generations of stars form out of the enriched ejecta of these stars (\citealt{Cottrell:1981}, \citealt{Gratton:2004}, \citealt{DErcole:2008}, \citealt{Decressin:2007a}). Other scenarios, involving accretion onto proto-planetary disks or very massive stars, have also been put forward as possible origins of abundance variations (\citealt{Bastian:2013}, \citealt{Bastian:2015}, \citealt{Denissenkov:2014}). However, as explored in depth by \cite{Bastian:2015} and \cite{Renzini:2015}, none of the currently proposed scenarios are able to fully explain the wide range of abundance patterns observed. A new, unknown mechanism may yet have to be devised to properly explain the range of observations.

Extension of observations to ultraviolet wavelengths has provided great gains in photometric evidence of multiple populations, especially with high precision space-based observations from the Hubble Space Telescope. The particular passbands F275W, F336W, and F438W disentangle multiple populations in globular clusters due to their sensitivity to helium and C, N, and O abundances. Specifically, the F275W filter contains an OH band, the F336W filter contains an NH band, and the F438W filter contains both CN and CH bands. These filters highlight distinctions among different metal contents correlated with helium and thus are able to visually show the chemical discontinuities between populations in color-magnitude space. Thus, a careful and thorough analysis of the color-magnitude diagram (CMD) allows a differentiation of the multiple populations of a cluster and their relative abundances.

We use the sensitive ultraviolet photometry from \cite{Piotto:2015} as well as visual photometry from \cite{Sarajedini:2007} to investigate the properties of 30 of the Galactic globular clusters. As presented in \cite{Stenning:2016} and \cite{Wagner-Kaiser:2016}, we employ a sophisticated Bayesian technique to fit theoretical models to the observed data. In the past, isochrones have primarily been fit to observations of star clusters by eye, trying different parameters and nudging isochrones in color-magnitude space until a solution looks good (\citealt{Jeffery:2016}). However, with our approach we are able to numerically determine the set of parameters for clusters using an in-depth grid of theoretical models (\citealt{von-Hippel:2006}, \citealt{De-Gennaro:2009}, \citealt{van-Dyk:2009}, \citealt{Stein:2013}, \citealt{Jeffery:2016}). We apply this method to 30 Galactic globular clusters, each fit with two dominant populations of stars, in order to better understand the relationships and origins of multiple populations.

In Section \ref{Data}, we discuss the photometry used to explore the multiple population phenomenon. In Section \ref{Methods}, the Bayesian framework is briefly summarized, with further details in \cite{Stenning:2016}. In Section \ref{Results}, our results are compared to previous studies and the primary results from our investigation are presented. We discuss formation scenarios and examine inconsistencies between the models and data in Section \ref{Discussion} and conclude in Section \ref{Conclusions}.


\section{Data}\label{Data}

HST Cycle 21 Program GO 13297 (PI: Piotto) observed 48 globular clusters, extending the wavelength coverage from the ACS Globular Treasury Survey (GO Cycle 14 Program 10775; PI: Sarajedini) into the ultraviolet with the F275W, F336W, and F438W filters. From these and nine previously observed clusters, we analyze a subset of 30. We provide a summary of the clusters analyzed in this work in Table \ref{clusterlist}.

\begin{figure*}
\includegraphics[width=0.95\textwidth]{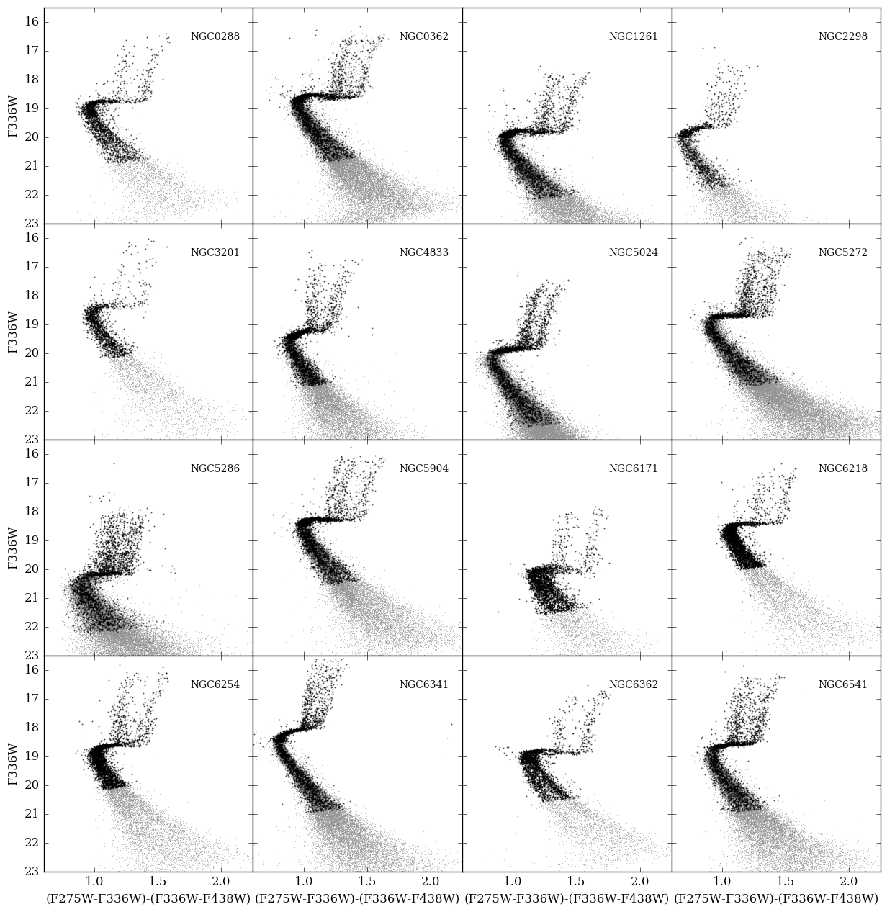}
\caption{Ultraviolet CMDs of clusters in our study. The color on the x-axis is chosen as a combination of the three ultraviolet filters to maximize the separation of the populations in the CMD. The clusters show two primary populations as seen in the red giant branches and main sequences (though further sub-populations may be present).}
\label{UVCMDs1}
\end{figure*}

\begin{figure*}
\includegraphics[width=0.95\textwidth]{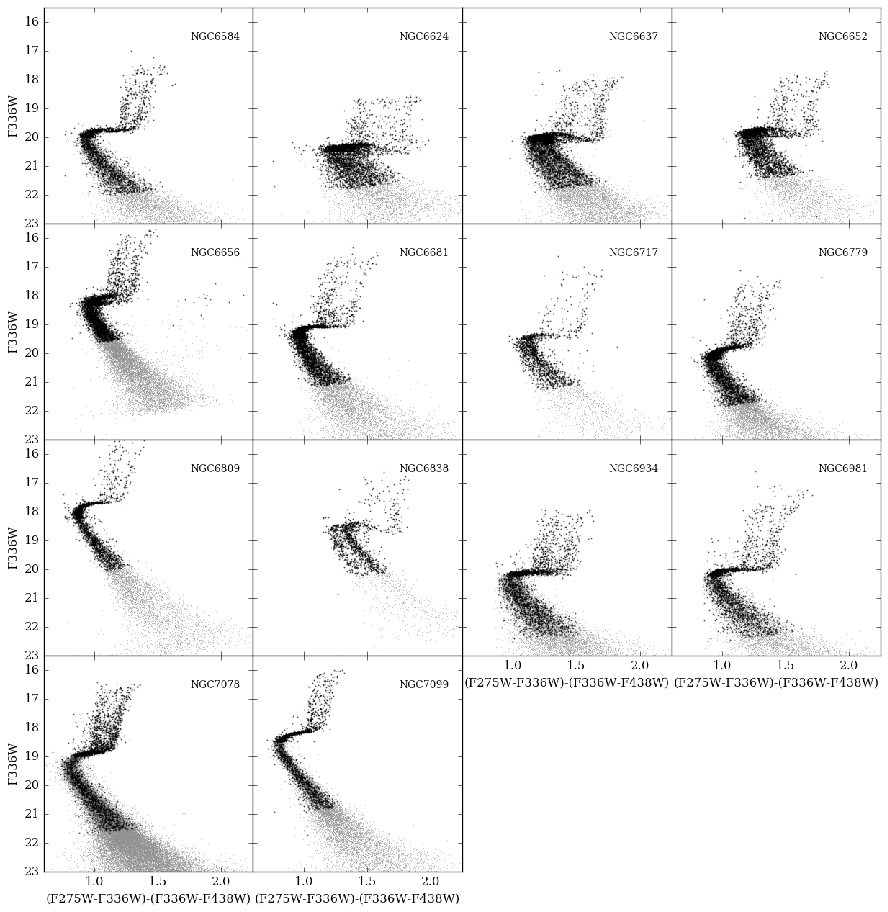}
\caption{Same as Figure \ref{UVCMDs1}, for the remainder of the clusters.}
\label{UVCMDs2}
\end{figure*}

Based on visual inspection of the CMDs, we choose the clusters for our sample because they appear to have two primary populations of stars, as shown in Figures \ref{UVCMDs1} and \ref{UVCMDs2}, an assumption that goes into our statistical model (as described in Section \ref{Methods})\footnote{Although likely harboring two primary populations, the MCMC chains for NGC6397, NGC6496 and NGC6535 did not converge under our current assumptions and were removed from our cluster sample.}. However, we note that additional sub-populations may be present in these clusters. In particular, NGC 362, NGC 5904, and NGC 6624 may have a less prominent third population. As our methodology is predicated on the presence of two populations, in these cases, our methodology will gravitate towards fitting the two photometrically dominant populations in the cluster. We do not expect to detect chemically differentiated sub-populations that are not photometrically distinct. We use photometry from the five available filters (F275W, F336W, F438W, F606W, F814W) in our analysis.

The current state of processing the HST ultraviolet observations is at the intermediate level photometry (see \citealt{Piotto:2015} for details). This provides a unified star list for the F275W, F336W, and F438W filters as well as the F606W and F814W filters from the ACS Globular Cluster Treasury Survey (\citealt{Sarajedini:2007}). We use differential reddening estimates for each star to correct for differential reddening in the clusters via a differential reddening map (\citealt{Milone:2012c, Piotto:2015}). Photometric errors for the three UV filters are estimated via RMS deviations from frame to frame. Uncertainties in the visual filters are derived from artificial star tests completed by the ACS Treasury Survey team. Pixel position errors are used as a probe of proper motion, and stars with large frame-to-frame position errors are removed in order to clean the photometry of the most obvious non-cluster stars. Photometric quality flags are used to remove stars with poor photometry (see \citealt{Piotto:2015} for further processing details).

We also remove horizontal branch (HB) stars from the photometry. Although the HB could provide valuable constraints, being particularly sensitive to changes in helium, HB models are not currently available in the DSED stellar evolution models for the HST UVIS filters. Alternate model sets either do not currently offer distinct variations in helium, lack equivalent evolutionary points needed for accurate model interpolation, or do not cover the necessary parameter space on the HB. We anticipate future development will capitalize on the information stored in the HB stars.

CMDs of the clusters are shown in Figures \ref{UVCMDs1} and \ref{UVCMDs2}, where the quantity plotted on the x-axis is a linear combination of the HST ultraviolet filters (specifically, (F275W--F336W) -- (F336W--F438W)) to visually accentuate the bi-modality of populations in these clusters (\citealt{Piotto:2015}).

\begin{table*}
\caption{Priors and Starting Values for Cluster Sample$^{a}$}
\centering
\begin{threeparttable}[b]
    \begin{tabular}{@{}ccccc@{}}
    \hline
 \textbf{Cluster} & \multicolumn{2}{c}{\textbf{Prior Distribution}}   & \textbf{Set Value} & \textbf{Starting Value} \\ \cline{2-3}
 \textbf{Name} & \textbf{Distance Modulus}  & \textbf{A$_{V}$}  & \textbf{[Fe/H]} & \textbf{Age$^b$ (Gyr)} \\  
\hline
NGC0288 	 & 14.84 $\pm$0.05 	 & 0.093 $\pm$0.03 	& -1.32 	  & 12.5 \\
NGC0362 	 & 14.83 $\pm$0.05 	 & 0.155 $\pm$0.05 	& -1.26 	 & 11.5 \\
NGC1261 	 & 16.09 $\pm$0.05 	& 0.031 $\pm$0.01 	& -1.27 	 & 11.5 \\
NGC2298 	 & 15.6 $\pm$0.05 	  & 0.434 $\pm$0.14 	& -1.97$^d$ 	 & 13.0 \\
NGC3201 	 & 14.2 $\pm$0.05 	 & 0.744 $\pm$0.25 	 & -1.56$^d$ 	 & 12.0 \\
NGC4833 	 & 15.08 $\pm$0.05 	 & 0.992 $\pm$0.33 	& -1.85 	  & 13.0 \\
NGC5024 	 & 16.32 $\pm$0.05 	& 0.062 $\pm$0.02 	  & -2.1 	 & 13.25 \\
NGC5272 	 & 15.07 $\pm$0.05 	  & 0.031 $\pm$0.01 	 & -1.524$^d$	& 12.5 \\
NGC5286 	 & 16.08 $\pm$0.05 	  & 0.744 $\pm$0.25 	 & -1.51$^d$ 	& 13.0 \\
NGC5904 	 & 14.46 $\pm$0.05 	  & 0.093 $\pm$0.03 	 & -1.26$^d$ 	& 12.25 \\
NGC6171 	 & 15.05 $\pm$0.05 	  & 1.023 $\pm$0.34 	 & -1.13$^d$ 	& 12.75 \\
NGC6218 	 & 14.01 $\pm$0.05 	  & 0.589 $\pm$0.2 	 & -1.50$^d$ 	& 13.25 \\
NGC6254 	 & 14.08 $\pm$0.05 	 & 0.868 $\pm$0.29 	 & -1.51$^d$ 	 & 13.0 \\
NGC6341 	 & 14.65 $\pm$0.05 	  & 0.062 $\pm$0.02 	 & -2.31 	& 13.25 \\
NGC6362 	 & 14.68 $\pm$0.05 	 & 0.279 $\pm$0.09 	 & -1.17$^d$ 	 & 12.5 \\
NGC6541 	 & 14.82 $\pm$0.05 	 & 0.434 $\pm$0.14 	 & -1.81 	 & 13.25 \\
NGC6584 	 & 15.96 $\pm$0.05 	 & 0.31 $\pm$0.1 	  & -1.5 	 & 12.25 \\
NGC6624 	 & 15.36 $\pm$0.05 	 & 0.868 $\pm$0.29 	 & -0.70$^d$ 	 & 13.0 \\
NGC6637 	 & 15.28 $\pm$0.05 	 & 0.558 $\pm$0.19 	 & -0.78$^d$ 	 & 12.5 \\
NGC6652 	 & 15.28 $\pm$0.05 	 & 0.279 $\pm$0.09 	 & -1.01$^d$ 	 & 13.25 \\
NGC6656 	 & 13.6 $\pm$0.05 	 & 1.054 $\pm$0.35 	 & -1.7 	 & 12.0$^c$ \\
NGC6681 	 & 14.99 $\pm$0.05 	& 0.217 $\pm$0.07 	  & -1.62 	 & 13.0 \\
NGC6717 	 & 14.94 $\pm$0.05 	 & 0.682 $\pm$0.23 	 & -1.26 	 & 13.0 \\
NGC6779 	 & 15.68 $\pm$0.05 	  & 0.806 $\pm$0.27 	 & -1.98 	& 13.5 \\
NGC6809 	 & 13.89 $\pm$0.05 	 & 0.248 $\pm$0.08 	 & -1.94 	 & 13.5 \\
NGC6838 	 & 13.8 $\pm$0.05 	 & 0.775 $\pm$0.26 	 & -0.78 	 & 12.5 \\
NGC6934 	 & 16.28 $\pm$0.05 	 & 0.31 $\pm$0.1 	 & -1.42$^d$ 	 & 12.0 \\
NGC6981 	 & 16.31 $\pm$0.05 	 & 0.155 $\pm$0.05 	 & -1.42 	 & 12.75 \\
NGC7078 	 & 15.39 $\pm$0.05 	 & 0.31 $\pm$0.1 	 & -2.37 	 & 13.25 \\
NGC7099 	 & 14.64 $\pm$0.05 	 & 0.093 $\pm$0.03 	 & -2.27 	 & 13.25 \\
\hline
    \end{tabular}
\begin{tablenotes}[b]
	\item $^a$ Results from \cite{Harris:2010} unless otherwise noted.
	\item $^b$ From \cite{Dotter:2010}.
	\item $^c$ No age estimate from \cite{Dotter:2010}; 12 Gyr is used as starting age.
	\item $^d$ Integrated spectroscopic metallicity from \cite{Schiavon:2005}.
\end{tablenotes}
\end{threeparttable}
\label{clusterlist}
\end{table*}

\section{Methods}\label{Methods}

We use a sophisticated Bayesian method to estimate a set of parameters describing globular clusters and to quantify uncertainty in those parameters. An objective and reproducible Bayesian approach is preferable to fitting by eye for several reasons. Our method addresses complex non-linear correlations between parameters by simultaneously sampling the joint posterior distribution of many parameters. We are able to incorporate error estimates on each observed photometric magnitude. Bayesian analyses also provide a posterior probability distribution for each parameter, which can tell us more about the likely values of the parameters than standard point estimates and error bars, especially when posterior distributions are substantially non-Gaussian.

Previously, the BASE-9 software suite has been used to fit single isochrones to stellar clusters, but recent work has extended the statistical framework to address the issue of two populations of stars in a single cluster (\citealt{Stenning:2016}, \citealt{Wagner-Kaiser:2016}). We use the newly extended BASE-9 software to sample the joint posterior distribution of age, extinction, distance to a cluster, helium values for populations of stars in the cluster, and the percentage of stars in the first population. A large grid of isochrones is used to interpolate between models and we are able to achieve high precision estimates for each parameter. The details and initial application of this methodology are established and explored in \cite{Stenning:2016} and \cite{Wagner-Kaiser:2016}. The software is available as open source code from GitHub (https://github.com/argiopetech/base/releases) and via executables through Amazon Web Services. Installation and instruction may be found in the User Manual for BASE-9 (\citealt{von-Hippel:2014}). Our methods mirror those of \cite{Stenning:2016} and \cite{Wagner-Kaiser:2016} and are summarized below.

The hierarchical model for the single population version of BASE-9 simultaneously samples cluster-level parameters (i.e., distance, extinction, age) and individual star-specific parameters (i.e., mass, binarity, cluster membership). To characterize two populations in globular clusters, we have adapted this model to include population level parameters of helium and population percentage (the proportion of cluster stars belonging to population A). The cluster-level parameters are shared among all cluster stars, while the population-level parameters are common to stars in a specific population in the cluster. Stellar-level parameters vary on a star-by-star basis.

Although binary stars are part of the BASE-9 statistical model, including them in our present analysis is computationally prohibitive. As globular clusters have low binary fractions, treating all stars as singletons should not significantly impact the final results. However, future code development will provide this ability at reasonable computational cost. We return to this point briefly in Section \ref{sec:prop}.

We use Dartmouth Stellar Evolution Database (DSED) models generated in the HST UVIS and ACS filters (\citealt{Dotter:2008}) over a range of ages (9 to 15 Gyr), metallicities (--2.5 to +0.5), and helium values ($\sim$0.23 to 0.40) as the theoretical isochrone grid over which we explore the parameter space. As presented in \cite{Stenning:2016}, we use a Markov chain Monte Carlo (MCMC) algorithm to explore the joint posterior distribution of the cluster-level and population-level parameters; we marginalize over (i.e. integrate out) the stellar-level parameters to reduce the level of multi-modality and nonlinearities that can frustrate MCMC convergence (see \citealt{Stein:2013} for additional discussion). Specifically, we implement an adaptive Metropolis (AM) algorithm that uses observed correlations in the MCMC chain to make sampling more efficient. A key benefit of the AM algorithm is that the user need only provide starting values for the chain; additional intervention and tuning is generally not required as adaptation is done automatically. After an initial burn-in period, which is typically 1000 iterations unless more are required, we run the AM MCMC chain for at least 10,000 additional iterations, using one chain per cluster due to computational limitations.

With the cleaned photometry discussed in Section \ref{Data}, and as in \cite{Wagner-Kaiser:2016}, we randomly select a subsample of $\lesssim$ 3000 stars, with half above the main sequence turnoff point (MSTOP) of the cluster and half below the MSTOP. If there are fewer than 1500 stars above the MSTOP, we match the number of stars above and below the MSTOP. This procedure is adopted to ensure a reasonable sample of stars on the sub-giant and red-giant branches of the CMD, where effects of multiple populations are primarily observed in these CMDs, while not having so many stars as to be computationally prohibitive.

Because we are performing a Bayesian analysis, we must specify prior distributions which summarize our knowledge regarding the model parameters before considering the current data. For distance and extinction, we use Gaussian prior distributions (truncated to be positive for extinction) with means and standard deviations set according to the published estimates and error bars in \cite{Harris:2010}. Generally speaking these priors are centered on published estimates but we use conservative prior dispersions (e.g., prior standard deviations) to reduce the influence of the priors on our final estimates. Specifically, we use a Gaussian standard deviation of 0.05 for distance modulus. For extinction, we conservatively use one third of the published value of the standard deviation as a prior standard deviation, assuming a \cite{Cardelli:1989} R$_{V}$ = 3.1 reddening law. The prior means and prior standard deviations are listed in Table \ref{clusterlist}. We choose to fix metallicity at reasonable values from the literature (\citealt{Harris:2010, Schiavon:2005}) in order to better constrain the helium abundances of the clusters. These fixed values are noted in Table \ref{clusterlist}. 

Because there is no useful prior information on absolute age, helium, or the percentage of stars in each population, we assume uniform priors on these parameters. For age, the prior is uniform from 1 to 15 Gyr. For the helium of population A, the prior is uniform from 0.15 to 0.35 and for population B the prior ranges from 0.15 to 0.45 (with the constraint that Y$_{B}$ must be greater than Y$_{A}$). The prior distribution for the population proportion is also uniform, over the range 0 to 1. We assume a \cite{Miller:1979} initial mass function as the prior distribution of individual stellar masses. 

We use a Gaussian model for the photometric magnitudes of cluster stars, with known (independent) measurement errors contained in the (diagonal) variance-covariance matrix. For field stars we use a simple model whereby the magnitudes are assumed to be uniformly distributed over the range of the data; this simple model is adequate for identifying field stars (see\citealt{Stenning:2016}. for a simulation study). We choose a prior value for the membership probability of 0.95, as we expect the cleaned photometry discussed in Section \ref{Data} to be largely free of field star contamination. The detailed mathematical description and integration of the field star model into the Bayesian framework is available in \cite{Stenning:2016}.

Starting values for the AM MCMC chain are chosen to be equivalent to the prior means for distance and extinction. For age, the value given in Table \ref{clusterlist} from \cite{Dotter:2010} serves as a starting point for sampling. For helium, we use starting points of 0.247 and 0.33, with a starting value of 0.5 for the proportion of cluster stars in population A. 

The photometry and errors for stars in multiple filters are input to BASE-9 along with the prior distribution information and model grid. The result of the Bayesian analysis is a correlated sample from the joint posterior of the cluster-level and population-level parameters. When presenting our results in the following sections, we provide the median of the posterior distribution and the 90\% Bayesian credible interval to represent statistical error. This statistical error represents the precision of the fit rather than the astronomical belief in the value. The uncertainties and assumptions of the isochrones propagated through the Bayesian analysis would provide a measure of the astronomical error, but these are not available at this time.

 
\section{Results}\label{Results}

In this section, we present the results of our analysis. Table \ref{clusterresults} provides the posterior medians and 90\% Bayesian credible intervals for age, distance, extinction, helium for each populations A and B, and the proportion (i.e., percentage of stars in population A). Throughout our analysis, we use these medians and intervals to represent the posterior distribution of the cluster and population parameters.

\begin{figure*} 
\includegraphics[width=0.95\textwidth]{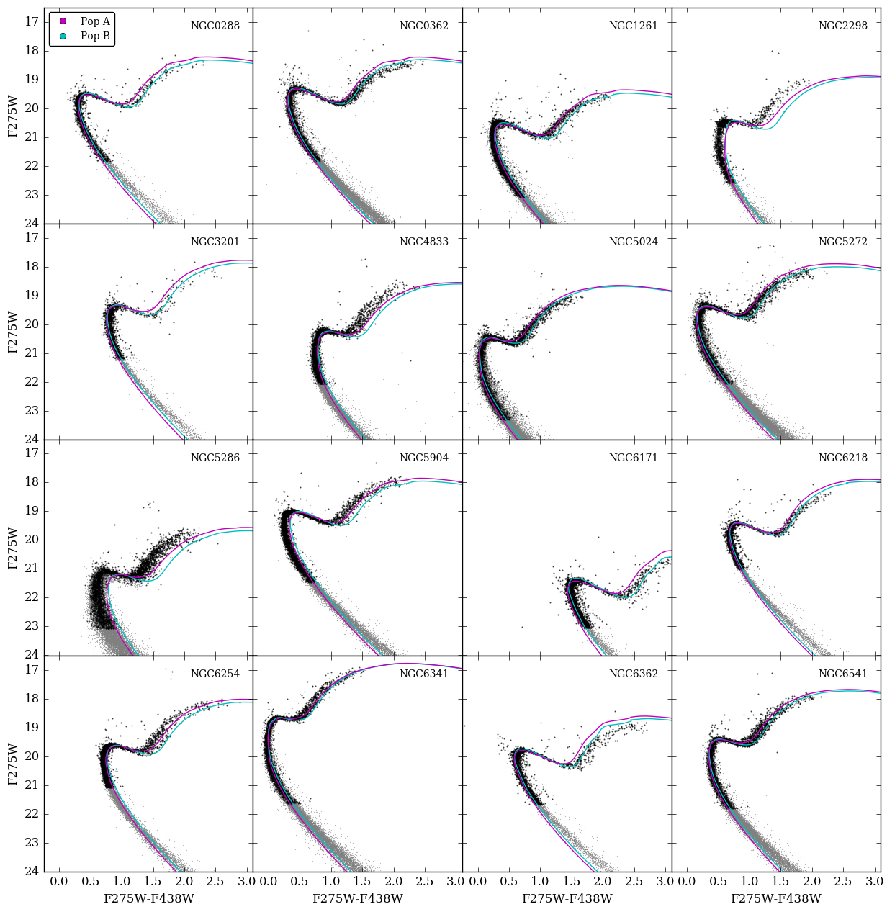}
\caption{The F275W--F438W CMD for 16 of the clusters in our sample. The gray points show the cleaned photometry and the black points show the randomly sampled subset of stars used in the Bayesian analysis. The isochrones are generated using the median of each posterior distribution, with cyan reflecting the helium fraction determined for population A and magenta representing population B.}
\label{CMDS1}
\end{figure*}

\begin{figure*} 
\includegraphics[width=0.95\textwidth]{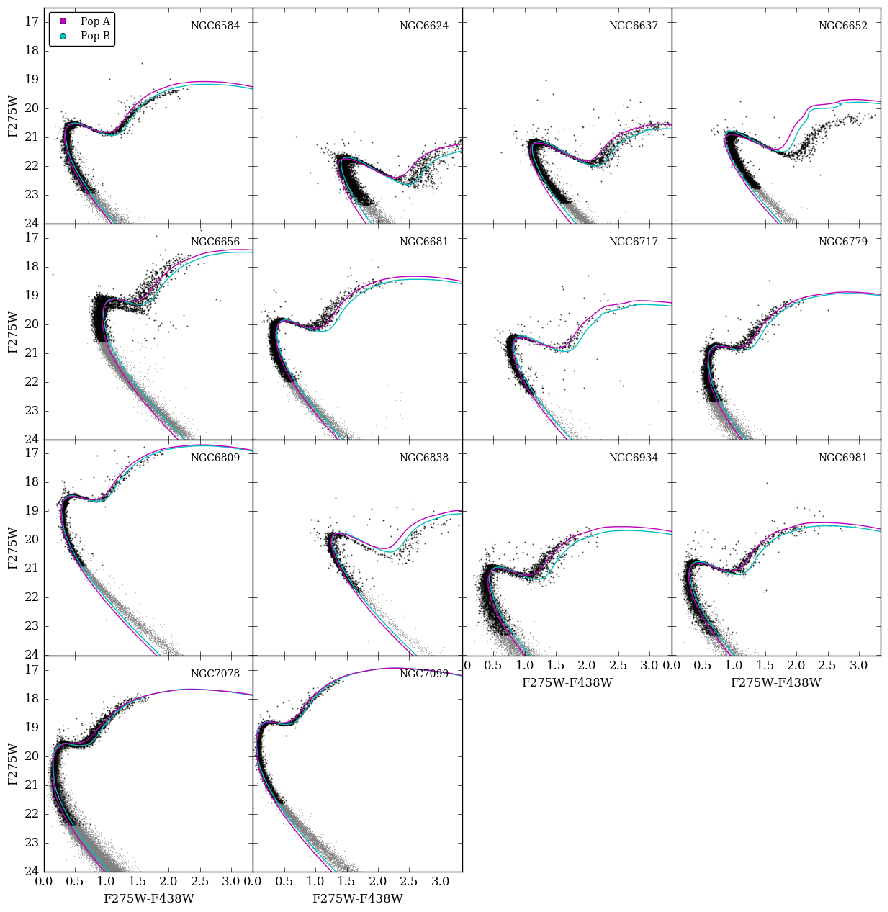}
\caption{Same as Figure \ref{CMDS1}, for the remaining clusters.}
\label{CMDS2}
\end{figure*}

In Figures \ref{CMDS1} and \ref{CMDS2}, our results are represented visually with a CMD based on F275W and F438W for each cluster. Each panel shows the cleaned photometry of a cluster in gray points, with the black points representing the selected subsample of stars used in the Bayesian analysis. The posterior medians presented in Table \ref{clusterresults} are used to generate an isochrone to represent each population, with the cyan isochrone representing the less enriched population A and the magenta representing the chemically processed population B. As discussed further in Section \ref{modelincon}, some CMDs appear to fit better in different color-magnitude combinations.

\renewcommand{\arraystretch}{2}
\begin{table*}
\centering
    \caption{Two-Population Bayesian Analysis}
    \scalebox{0.9}{
    \begin{tabular}{@{}cccccccccc@{}}
    \hline
 \textbf{Name} &  \textbf{[Fe/H]} & \textbf{log(Age)}  & \textbf{Age (Gyr)}  & \textbf{Distance Modulus}  & \textbf{A$_{V}$}  & \textbf{Y$_{A}$}	& \textbf{Y$_{B}$} & \textbf{$\Delta$Y}	& \textbf{Proportion}\\    \hline
NGC0288 	 & -1.32 	 & 10.043$^{0.001}_{0.002}$ 	 & 11.033$^{0.036}_{0.042}$ 	 & 15.069$^{0.004}_{0.004}$ 	 & 0.140$^{0.002}_{0.001}$ 	 & 0.272$^{0.002}_{0.001}$ 	 & 0.328$^{0.002}_{0.001}$ 	 & 0.055$^{0.002}_{0.002}$ 	 & 0.553$^{0.031}_{0.031}$ 	\\
NGC0362 	 & -1.26 	 & 10.000$^{0.001}_{0.001}$ 	 & 9.992$^{0.024}_{0.025}$ 	 & 14.984$^{0.003}_{0.003}$ 	 & 0.128$^{0.001}_{0.001}$ 	 & 0.269$^{0.001}_{0.001}$ 	 & 0.314$^{0.001}_{0.001}$ 	 & 0.045$^{0.002}_{0.002}$ 	 & 0.487$^{0.022}_{0.025}$ 	\\
NGC1261 	 & -1.27 	 & 10.006$^{0.001}_{0.001}$ 	 & 10.142$^{0.025}_{0.027}$ 	 & 16.222$^{0.003}_{0.003}$ 	 & 0.088$^{0.001}_{0.001}$ 	 & 0.274$^{0.001}_{0.002}$ 	 & 0.333$^{0.001}_{0.001}$ 	 & 0.059$^{0.002}_{0.002}$ 	 & 0.374$^{0.021}_{0.019}$ 	\\
NGC2298 	 & -1.97 	 & 10.017$^{0.002}_{0.002}$ 	 & 10.401$^{0.049}_{0.047}$ 	 & 15.862$^{0.006}_{0.006}$ 	 & 0.885$^{0.002}_{0.002}$ 	 & 0.266$^{0.005}_{0.004}$ 	 & 0.373$^{0.004}_{0.003}$ 	 & 0.106$^{0.006}_{0.005}$ 	 & 0.318$^{0.035}_{0.033}$ 	\\
NGC3201 	 & -1.56 	 & 10.064$^{0.002}_{0.002}$ 	 & 11.585$^{0.040}_{0.042}$ 	 & 14.244$^{0.004}_{0.004}$ 	 & 0.904$^{0.001}_{0.001}$ 	 & 0.286$^{0.002}_{0.002}$ 	 & 0.346$^{0.002}_{0.002}$ 	 & 0.060$^{0.002}_{0.002}$ 	 & 0.421$^{0.026}_{0.033}$ 	\\
NGC4833 	 & -1.85 	 & 10.072$^{0.002}_{0.002}$ 	 & 11.795$^{0.058}_{0.068}$ 	 & 15.172$^{0.007}_{0.006}$ 	 & 1.052$^{0.002}_{0.002}$ 	 & 0.311$^{0.003}_{0.005}$ 	 & 0.385$^{0.003}_{0.004}$ 	 & 0.073$^{0.004}_{0.006}$ 	 & 0.513$^{0.039}_{0.038}$ 	\\
NGC5024 	 & -2.10 	 & 10.108$^{0.001}_{0.001}$ 	 & 12.821$^{0.035}_{0.034}$ 	 & 16.505$^{0.003}_{0.003}$ 	 & 0.094$^{0.001}_{0.001}$ 	 & 0.257$^{0.002}_{0.002}$ 	 & 0.336$^{0.001}_{0.001}$ 	 & 0.078$^{0.002}_{0.002}$ 	 & 0.436$^{0.023}_{0.023}$ 	\\
NGC5272 	 & -1.52 	 & 10.053$^{0.001}_{0.001}$ 	 & 11.290$^{0.026}_{0.026}$ 	 & 15.201$^{0.003}_{0.003}$ 	 & 0.085$^{0.001}_{0.001}$ 	 & 0.260$^{0.001}_{0.001}$ 	 & 0.318$^{0.001}_{0.001}$ 	 & 0.058$^{0.002}_{0.002}$ 	 & 0.404$^{0.024}_{0.024}$ 	\\
NGC5286 	 & -1.51 	 & 10.041$^{0.001}_{0.002}$ 	 & 10.978$^{0.026}_{0.049}$ 	 & 16.068$^{0.005}_{0.004}$ 	 & 0.948$^{0.002}_{0.002}$ 	 & 0.345$^{0.003}_{0.003}$ 	 & 0.434$^{0.003}_{0.003}$ 	 & 0.089$^{0.004}_{0.004}$ 	 & 0.479$^{0.027}_{0.028}$ 	\\
NGC5904 	 & -1.26 	 & 9.992$^{0.001}_{0.001}$ 	 & 9.817$^{0.024}_{0.025}$ 	 & 14.618$^{0.003}_{0.003}$ 	 & 0.194$^{0.001}_{0.001}$ 	 & 0.300$^{0.001}_{0.001}$ 	 & 0.354$^{0.001}_{0.001}$ 	 & 0.054$^{0.002}_{0.002}$ 	 & 0.503$^{0.022}_{0.022}$ 	\\
NGC6171 	 & -1.13 	 & 10.056$^{0.003}_{0.002}$ 	 & 11.371$^{0.071}_{0.059}$ 	 & 15.248$^{0.010}_{0.010}$ 	 & 1.459$^{0.002}_{0.003}$ 	 & 0.230$^{0.004}_{0.003}$ 	 & 0.306$^{0.003}_{0.004}$ 	 & 0.076$^{0.005}_{0.005}$ 	 & 0.343$^{0.019}_{0.019}$ 	\\
NGC6218 	 & -1.50 	 & 10.110$^{0.001}_{0.001}$ 	 & 12.877$^{0.022}_{0.025}$ 	 & 14.340$^{0.002}_{0.002}$ 	 & 0.695$^{0.001}_{0.001}$ 	 & 0.235$^{0.001}_{0.001}$ 	 & 0.278$^{0.001}_{0.001}$ 	 & 0.043$^{0.002}_{0.002}$ 	 & 0.357$^{0.026}_{0.023}$ 	\\
NGC6254 	 & -1.51 	 & 10.079$^{0.001}_{0.000}$ 	 & 12.007$^{0.015}_{0.010}$ 	 & 14.477$^{0.002}_{0.001}$ 	 & 0.864$^{0.001}_{0.001}$ 	 & 0.295$^{0.002}_{0.002}$ 	 & 0.355$^{0.001}_{0.001}$ 	 & 0.060$^{0.002}_{0.002}$ 	 & 0.471$^{0.021}_{0.020}$ 	\\
NGC6341 	 & -2.31 	 & 10.115$^{0.001}_{0.001}$ 	 & 13.029$^{0.019}_{0.019}$ 	 & 14.769$^{0.002}_{0.002}$ 	 & 0.095$^{0.001}_{0.001}$ 	 & 0.314$^{0.001}_{0.002}$ 	 & 0.381$^{0.001}_{0.001}$ 	 & 0.067$^{0.002}_{0.002}$ 	 & 0.557$^{0.026}_{0.025}$ 	\\
NGC6362 	 & -1.17 	 & 10.108$^{0.001}_{0.001}$ 	 & 12.818$^{0.033}_{0.039}$ 	 & 14.823$^{0.003}_{0.003}$ 	 & 0.268$^{0.001}_{0.001}$ 	 & 0.220$^{0.001}_{0.001}$ 	 & 0.265$^{0.001}_{0.001}$ 	 & 0.045$^{0.002}_{0.002}$ 	 & 0.385$^{0.026}_{0.024}$ 	\\
NGC6541 	 & -1.81 	 & 10.104$^{0.001}_{0.001}$ 	 & 12.700$^{0.029}_{0.033}$ 	 & 14.879$^{0.003}_{0.003}$ 	 & 0.429$^{0.001}_{0.001}$ 	 & 0.297$^{0.001}_{0.001}$ 	 & 0.349$^{0.001}_{0.001}$ 	 & 0.052$^{0.002}_{0.002}$ 	 & 0.502$^{0.024}_{0.025}$ 	\\
NGC6584 	 & -1.50 	 & 10.054$^{0.001}_{0.001}$ 	 & 11.329$^{0.032}_{0.030}$ 	 & 16.077$^{0.003}_{0.003}$ 	 & 0.319$^{0.001}_{0.001}$ 	 & 0.260$^{0.002}_{0.002}$ 	 & 0.311$^{0.001}_{0.001}$ 	 & 0.051$^{0.002}_{0.002}$ 	 & 0.431$^{0.031}_{0.029}$ 	\\
NGC6624 	 & -0.70 	 & 9.957$^{0.003}_{0.002}$ 	 & 9.065$^{0.062}_{0.046}$ 	 & 15.793$^{0.006}_{0.007}$ 	 & 0.969$^{0.003}_{0.003}$ 	 & 0.265$^{0.001}_{0.002}$ 	 & 0.343$^{0.002}_{0.002}$ 	 & 0.077$^{0.002}_{0.003}$ 	 & 0.677$^{0.021}_{0.016}$ 	\\
NGC6637 	 & -0.78 	 & 9.983$^{0.003}_{0.002}$ 	 & 9.622$^{0.063}_{0.050}$ 	 & 15.618$^{0.006}_{0.007}$ 	 & 0.665$^{0.002}_{0.003}$ 	 & 0.265$^{0.001}_{0.001}$ 	 & 0.330$^{0.000}_{0.000}$ 	 & 0.065$^{0.001}_{0.001}$ 	 & 0.660$^{0.017}_{0.019}$ 	\\
NGC6652 	 & -1.10 	 & 10.176$^{0.000}_{0.000}$ 	 & 14.999$^{0.001}_{0.002}$ 	 & 15.432$^{0.002}_{0.002}$ 	 & 0.452$^{0.001}_{0.001}$ 	 & 0.178$^{0.001}_{0.002}$ 	 & 0.230$^{0.002}_{0.003}$ 	 & 0.052$^{0.002}_{0.003}$ 	 & 0.560$^{0.027}_{0.023}$ 	\\
NGC6656 	 & -1.70 	 & 10.078$^{0.001}_{0.002}$ 	 & 11.971$^{0.030}_{0.052}$ 	 & 13.732$^{0.006}_{0.005}$ 	 & 1.210$^{0.002}_{0.002}$ 	 & 0.327$^{0.002}_{0.003}$ 	 & 0.417$^{0.003}_{0.003}$ 	 & 0.090$^{0.004}_{0.005}$ 	 & 0.446$^{0.029}_{0.031}$ 	\\
NGC6681 	 & -1.62 	 & 10.072$^{0.001}_{0.001}$ 	 & 11.809$^{0.035}_{0.033}$ 	 & 15.364$^{0.003}_{0.003}$ 	 & 0.393$^{0.001}_{0.001}$ 	 & 0.248$^{0.002}_{0.001}$ 	 & 0.312$^{0.001}_{0.001}$ 	 & 0.063$^{0.002}_{0.002}$ 	 & 0.494$^{0.023}_{0.023}$ 	\\
NGC6717 	 & -1.26 	 & 10.062$^{0.002}_{0.001}$ 	 & 11.526$^{0.047}_{0.030}$ 	 & 15.226$^{0.003}_{0.004}$ 	 & 0.728$^{0.002}_{0.002}$ 	 & 0.250$^{0.002}_{0.002}$ 	 & 0.316$^{0.002}_{0.002}$ 	 & 0.066$^{0.003}_{0.003}$ 	 & 0.487$^{0.034}_{0.033}$ 	\\
NGC6779 	 & -1.98 	 & 10.130$^{0.001}_{0.001}$ 	 & 13.488$^{0.018}_{0.028}$ 	 & 15.861$^{0.002}_{0.002}$ 	 & 0.799$^{0.001}_{0.001}$ 	 & 0.283$^{0.002}_{0.002}$ 	 & 0.365$^{0.002}_{0.002}$ 	 & 0.082$^{0.003}_{0.003}$ 	 & 0.435$^{0.023}_{0.023}$ 	\\
NGC6809 	 & -1.94 	 & 10.107$^{0.001}_{0.001}$ 	 & 12.782$^{0.034}_{0.034}$ 	 & 14.068$^{0.003}_{0.003}$ 	 & 0.392$^{0.001}_{0.001}$ 	 & 0.285$^{0.002}_{0.002}$ 	 & 0.335$^{0.001}_{0.001}$ 	 & 0.050$^{0.002}_{0.002}$ 	 & 0.497$^{0.033}_{0.031}$ 	\\
NGC6838 	 & -0.78 	 & 10.013$^{0.008}_{0.003}$ 	 & 10.304$^{0.198}_{0.062}$ 	 & 13.853$^{0.008}_{0.023}$ 	 & 0.839$^{0.003}_{0.006}$ 	 & 0.301$^{0.003}_{0.003}$ 	 & 0.341$^{0.004}_{0.002}$ 	 & 0.040$^{0.005}_{0.003}$ 	 & 0.400$^{0.033}_{0.036}$ 	\\
NGC6934 	 & -1.42 	 & 10.051$^{0.001}_{0.001}$ 	 & 11.256$^{0.035}_{0.037}$ 	 & 16.345$^{0.004}_{0.003}$ 	 & 0.372$^{0.002}_{0.002}$ 	 & 0.271$^{0.008}_{0.002}$ 	 & 0.350$^{0.002}_{0.001}$ 	 & 0.079$^{0.008}_{0.003}$ 	 & 0.365$^{0.028}_{0.022}$ 	\\
NGC6981 	 & -1.42 	 & 10.046$^{0.001}_{0.001}$ 	 & 11.105$^{0.030}_{0.033}$ 	 & 16.386$^{0.003}_{0.003}$ 	 & 0.187$^{0.001}_{0.001}$ 	 & 0.266$^{0.002}_{0.002}$ 	 & 0.330$^{0.001}_{0.000}$ 	 & 0.064$^{0.002}_{0.002}$ 	 & 0.410$^{0.026}_{0.023}$ 	\\
NGC7078 	 & -2.37 	 & 10.109$^{0.001}_{0.001}$ 	 & 12.856$^{0.027}_{0.028}$ 	 & 15.472$^{0.003}_{0.003}$ 	 & 0.317$^{0.001}_{0.001}$ 	 & 0.326$^{0.002}_{0.002}$ 	 & 0.410$^{0.002}_{0.002}$ 	 & 0.084$^{0.003}_{0.003}$ 	 & 0.525$^{0.025}_{0.023}$ 	\\
NGC7099 	 & -2.27 	 & 10.108$^{0.001}_{0.001}$ 	 & 12.811$^{0.029}_{0.026}$ 	 & 14.827$^{0.003}_{0.003}$ 	 & 0.171$^{0.001}_{0.001}$ 	 & 0.316$^{0.001}_{0.001}$ 	 & 0.377$^{0.002}_{0.002}$ 	 & 0.061$^{0.002}_{0.002}$ 	 & 0.610$^{0.026}_{0.028}$ 	\\
\hline
    \end{tabular}}
   \label{clusterresults}
\end{table*}
\renewcommand{\arraystretch}{1}

\subsection{Comparisons}\label{Comparisons}

\begin{figure*} 
\includegraphics[width=0.95\textwidth]{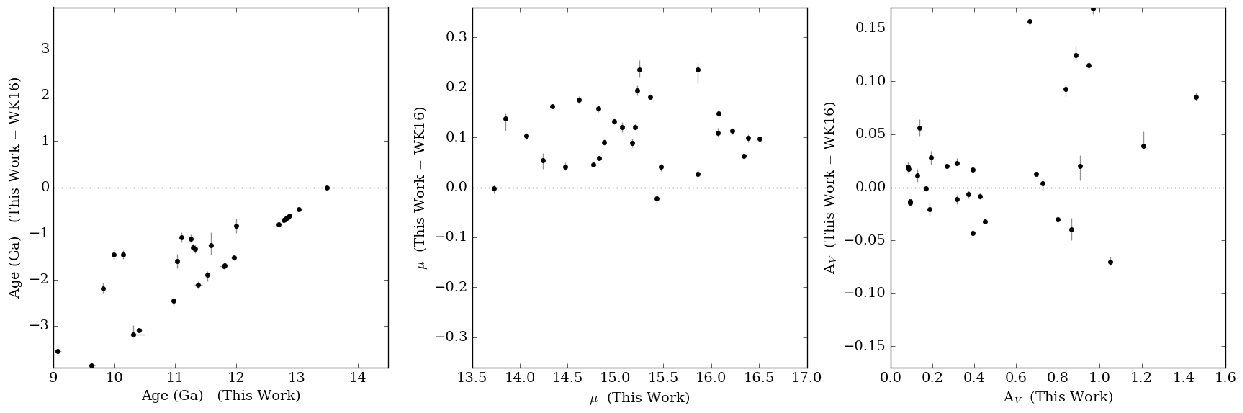}
\caption{This figure compares the results from this paper to the results from \protect\cite{Wagner-Kaiser:2016b} by plotting the difference vs the value from this paper, where the difference is calculated by subtracting the result from \protect\cite{Wagner-Kaiser:2016b} from the results herein. We compare the median of the posterior for age (left), distance modulus (middle), and extinction (right), with errors bars representing the 90\protect\% Bayesian credible intervals. We find generally younger ages than the single population analysis, which has a set age limit of 13.5 Gyr. Slight offsets in distance and absorption are observed.}
\label{Comp_plot}
\end{figure*}

We compare the ages, distances, and extinctions from the two-population analysis of the clusters from this work to the single population analysis from \cite{Wagner-Kaiser:2016b}. As shown in Figure \ref{Comp_plot}, we see offsets between the two analyses. Unlike in \cite{Wagner-Kaiser:2016b} where a single population and a single helium abundance are assumed, the current work operates with two separate populations that are distinct in helium. Analyzed with two populations, we find the clusters tend to be younger and at marginally greater distances than in the single population analysis of \cite{Wagner-Kaiser:2016b}. The additional information provided by the three ultraviolet filters necessarily affects the resulting posterior distribution for cluster and population parameters (\citealt{Hills:2015}).

In Table \ref{medcomp}, we present the offsets in the error-weighted medians\footnote{Weighted median = $\frac{\Sigma_{i=1}^{n}w_{i}X_{i}}{\Sigma_{i=1}^{n}w_{i}}$} with standard deviations in comparison to \cite{Wagner-Kaiser:2016b}, \cite{Dotter:2010}, and \cite{Harris:2010}. While most of the standard deviations are greater than the offsets, comparisons to  \cite{Dotter:2010} and \cite{Harris:2010} suggest that including the ultraviolet observations leads to younger ages. Again, our results give slightly larger distance moduli and marginally greater absorptions than previously published studies with our five-filter photometry.

\begin{table*}
\caption{Comparison to Previous Studies$^{1}$}
\centering
\begin{threeparttable}[b]
    \begin{tabular}{@{}cccc@{}}
    \hline
 \textbf{Parameter} &  \textbf{\cite{Wagner-Kaiser:2016b}} & \textbf{\cite{Dotter:2010}}  & \textbf{\cite{Harris:2010}} \\  
\hline
Age (Gyr) 	 & 	-1.861 $\pm$ 1.036	&	-1.092 $\pm$ 1.065	& --- 	\\
A$_{V}$ 	 &	0.020 $\pm$ 0.055	&	---				&	0.063 $\pm$ 0.110	\\
Distance 	 &	0.135 $\pm$ 0.091	&	---				&	0.138  $\pm$ 0.108	\\
\hline
    \end{tabular}
\begin{tablenotes}[b]
	\item $^1$ Median, error-weighted differences are calculated as This Work -- Referenced Work.
\end{tablenotes}
\end{threeparttable}
\label{medcomp}
\end{table*}

\renewcommand{\arraystretch}{1.15}
\begin{table*}
\caption{Comparison to Literature Values}
\centering
\begin{threeparttable}[b]
\scalebox{0.75}{
    \begin{tabular}{@{}llllll@{}}
    \hline
 \textbf{Cluster} &  \textbf{Y$_{A}$} & \textbf{Y$_{B}$}  & \textbf{$\Delta$Y}  & \textbf{Reference} \\  
\hline
NGC0288	& 	--- 	&	---	&	0.011 to 0.016; average: 0.013 $\pm$ 0.001	& \cite{Piotto:2013}   \\
		& 	0.28 &	0.292 &	0.012		&  \cite{Gratton:2010}$^3$   \\
		&	---	&	---	&	0.03			& \cite{Roh:2011} \\
		& 0.272$^{0.002}_{0.001}$ 	 & 0.328$^{0.002}_{0.001}$ 	 & 0.055$^{0.002}_{0.002}$   & This Work   \\
\hline
NGC0362	&	0.243 &	0.289 &	0.046		& \cite{Gratton:2010}$^3$   \\
		& 0.269$^{0.001}_{0.001}$ 	 & 0.314$^{0.001}_{0.001}$ 	 & 0.045$^{0.002}_{0.002}$   & This Work   \\
\hline
NGC1261	&	0.244 &	0.297 &	0.053		& \cite{Gratton:2010}$^3$   \\
		& 0.274$^{0.001}_{0.002}$ 	 & 0.333$^{0.001}_{0.001}$ 	 & 0.059$^{0.002}_{0.002}$ 	 & This Work   \\
\hline
NGC3201	&	0.253 &	0.278 &	0.025		& \cite{Gratton:2010}$^3$   \\
		& 0.286$^{0.002}_{0.002}$ 	 & 0.346$^{0.002}_{0.002}$ 	 & 0.060$^{0.002}_{0.002}$ 	 & This Work   \\
\hline
NGC5024	&	---	&	0.42$^1$ 	&	---		& \cite{Caloi:2011}   \\
		& 0.241	& 0.241		& 0.0			& \cite{Gratton:2010}$^3$   \\
		& 0.24	& 0.27-0.29 & 	0.03 to 0.05	& \cite{DAntona:2008}   \\
		& 0.257$^{0.002}_{0.002}$ 	 & 0.336$^{0.001}_{0.001}$ 	 & 0.078$^{0.002}_{0.002}$	 & This Work   \\
\hline
NGC5272	&	---	&	---	& 0 to 0.02			& \cite{Catelan:2009}, \cite{Valcarce:2010}, \\
		&		&		&				& \cite{Dalessandro:2013}   \\
		& 0.24	& 0.26 to 0.28 & 0.02 to 0.04	& \cite{Caloi:2008}   \\
		& 0.249	& 0.272	& 0.023			& \cite{Gratton:2010}$^3$   \\
		& 0.260$^{0.001}_{0.001}$ 	 & 0.318$^{0.001}_{0.001}$ 	 & 0.058$^{0.002}_{0.002}$  	 & This Work   \\
\hline
NGC5904	& 0.30$^2$& ---		&	---			& \cite{Khamidullina:2014}   \\
		& 0.25$^2$& ---		&	---			& \cite{vandenBerg:2013}   \\
		& 0.262	& 0.293	& 0.031			& \cite{Gratton:2010}$^3$   \\
		& 0.300$^{0.001}_{0.001}$ 	 & 0.354$^{0.001}_{0.001}$ 	 & 0.054$^{0.002}_{0.002}$  	 & This Work   \\
\hline
NGC6218	& 0.25	& 0.30	& 0.05			& \cite{Carretta:2007}   \\
		& 0.270	& 0.290	& 0.02			& \cite{Gratton:2010}$^3$   \\
		& 0.235$^{0.001}_{0.001}$ 	 & 0.278$^{0.001}_{0.001}$ 	 & 0.043$^{0.002}_{0.002}$  	 & This Work   \\
\hline
NGC6254	& 0.3	$^2$& ---		& ---				& \cite{Khamidullina:2014}   \\
		& 0.25$^2$& ---		& ---				& \cite{vandenBerg:2013}   \\
		& 0.287	& 0.327	& 0.040			& \cite{Gratton:2010}$^3$   \\
		& 0.295$^{0.002}_{0.002}$ 	 & 0.355$^{0.001}_{0.001}$ 	 & 0.060$^{0.002}_{0.002}$ 	 & This Work   \\
\hline
NGC6362 & 0.292$\pm$0.002$^2$& ---	& ---		& \cite{Olech:2001}   \\
		& 0.237	& 0.260	& 0.023			& \cite{Gratton:2010}$^3$   \\
		& 0.220$^{0.001}_{0.001}$ 	 & 0.265$^{0.001}_{0.001}$ 	 & 0.045$^{0.002}_{0.002}$ 	 & This Work   \\
\hline
NGC6656 & 0.252	& 0.269	& 0.017			& \cite{Gratton:2010}$^3$   \\
		& ---		& ---		& 0.09			& \cite{Joo:2013} \\
		& 0.327$^{0.002}_{0.003}$ 	 & 0.417$^{0.003}_{0.003}$ 	 & 0.090$^{0.004}_{0.005}$ 	 & This Work   \\
\hline
NGC6779	& 0.23$^2$& ---		& ---				& \cite{Khamidullina:2014}   \\
		& 0.25$^2$& ---		& ---				& \cite{vandenBerg:2013}   \\
		& 0.247	& 0.271	& 0.024			& \cite{Gratton:2010}$^3$   \\
		& 0.283$^{0.002}_{0.002}$ 	 & 0.365$^{0.002}_{0.002}$ 	 & 0.082$^{0.003}_{0.003}$  	 & This Work   \\
\hline
NGC6809	& 0.274$\pm$0.016$^2$ & ---	& ---				& \cite{Vargas-Alvarez:2007}   \\
		& 0.25	& 0.275			& 0.025			& \cite{Gratton:2010}$^3$   \\
		& 0.285$^{0.002}_{0.002}$ 	 & 0.335$^{0.001}_{0.001}$ 	 & 0.050$^{0.002}_{0.002}$	 & This Work   \\
\hline
NGC6934	& 0.27$^2$ & ---	& ---				& \cite{Kaluzny:2001}   \\
		& 0.238	& 0.283	& 0.045			& \cite{Gratton:2010}$^3$   \\
		 & 0.271$^{0.008}_{0.002}$ 	 & 0.350$^{0.002}_{0.001}$ 	 & 0.079$^{0.008}_{0.003}$ 	 & This Work   \\
\hline
NGC7078 & 0.232	& 0.305	& 0.073			& \cite{Gratton:2010}$^3$   \\
		& 0.326$^{0.002}_{0.002}$ 	 & 0.410$^{0.002}_{0.002}$ 	 & 0.084$^{0.003}_{0.003}$  	 & This Work   \\
\hline
NGC7099 & 0.245	& 0.249	& 0.004			& \cite{Gratton:2010}$^3$   \\
		& 0.316$^{0.001}_{0.001}$ 	 & 0.377$^{0.002}_{0.002}$ 	 & 0.061$^{0.002}_{0.002}$  	& This Work   \\
\hline
    \end{tabular}}
\begin{tablenotes}[b]
	\item $^1$ Upper limit
	\item $^2$ Overall Y estimate for the cluster
	\item $^3$ Y$_{A}$ is characterized by Y$_{med}$ from \cite{Gratton:2010}, which may overestimate the minimum helium fraction and underestimate $\Delta$Y in the cluster
\end{tablenotes}
\end{threeparttable}
\label{litcomp}
\end{table*}
\renewcommand{\arraystretch}{1}

In Table \ref{litcomp}, for 16 clusters, we compare our values to previous studies that have examined the helium content in these clusters. In many cases, our values are commensurate with previous estimates. However, a decrease in light element abundance (i.e.: [$\alpha$/Fe]) mimics an increase in the helium fraction in the morphology of an isochrone. Because we do not take light element abundance variations into account in this analysis, it is possible that some of the variation in helium ($\Delta$Y) is attributable to changes in light elements among the populations. We expect that our results estimate the upper limit of helium variation in each cluster.

We find greater spreads in $\Delta$Y for the clusters also studied by \cite{Gratton:2010}, who obtain estimates of helium using the horizontal branch and the \emph{R}-parameter (\cite{Salaris:2004}). However, they note that their determinations of Y$_{\text{med}}$ (which we list under the Y$_{A}$ column) may be overestimated. This would mean their determinations of $\Delta$Y could be underestimated and perhaps accounts for the difference in values between our study and theirs. In comparison to prior studies of NGC 3201, NGC 6779, NGC 6934, and NGC 7099, our determination of the spread of helium within the cluster tend to be greater by more than 0.03 dex. Our estimates are comparable within 0.03 dex of previously published $\Delta$Y estimates for other clusters. Considering the current variety of methodologies and theoretical models used to obtain estimates of $\Delta$Y, we believe such discrepancies from study to study are not unreasonable. However, we present here the largest sample to date of consistently determined characterizations of two populations in Galactic clusters via isochrone fitting.

For several published studies, only one value of helium is determined for the cluster, which is often bracketed by our own helium determinations of each population. However, we note many studies conducted by others lack error bars, making it difficult to determine how many of our estimates are truly consistent with previous studies.

\subsection{Helium}

\begin{figure} 
\includegraphics[width=0.5\textwidth]{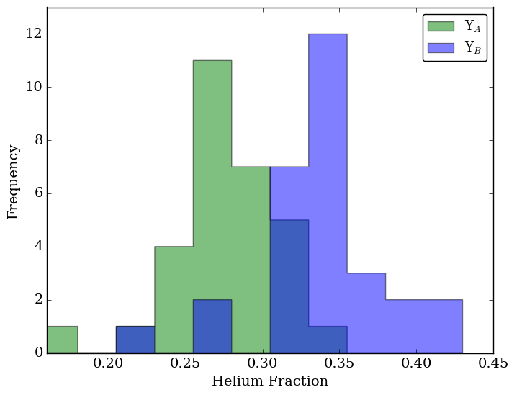}
\caption{A histogram of the distribution of the median of the posterior helium fractions of each population, with Y$_{A}$ in green and Y$_{B}$ in blue. The distribution of helium fractions for population A in the clusters has a peak around 0.27 with a range from $\sim$0.17 to 0.34. The helium peaks at $\sim$0.34 for population B and ranges from 0.23 to 0.43.}
\label{Y_hist}
\end{figure}

In Figure \ref{Y_hist}, we present the medians of the helium fraction posteriors among the studied clusters. Both distributions are relatively normally distributed. Values for the helium fraction of population A range from about 0.17 to 0.33 with a peak at 0.24. For population B, the helium fraction ranges from $\sim$0.22 to 0.39, with a peak around 0.31.

\begin{figure} 
\includegraphics[width=0.5\textwidth]{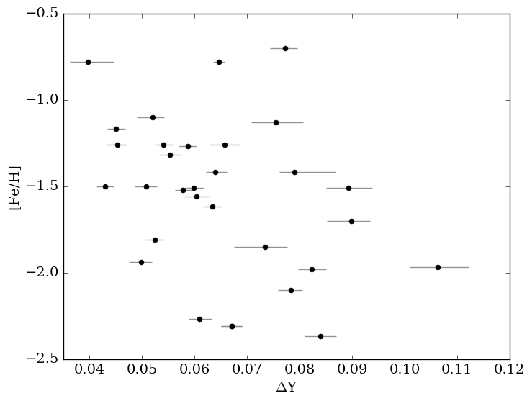}
\caption{The relationship between metallicity and $\Delta$Y, the median posterior of the difference between the helium of populations A and B in each of the clusters.}
\label{FeH_DY}
\end{figure}

As seen in Figure \ref{FeH_DY}, the $\Delta$Y values tend to fall into the relatively narrow range of $\sim$0.04 to 0.11, with a median around 0.063. This follows the results of \cite{Bragaglia:2010}, who suggested that the range of helium values in clusters is likely to be around 0.05 to 0.10. To the eye, it appears as though the more metal-poor clusters may have marginally larger spreads in helium than the metal-rich clusters. However, there is no statistically significant relationship between the spread of helium and the metallicity of the cluster.

\subsection{Mass}\label{sec:mass}


\cite{Gratton:2010} showed the spread of the Na-O anti correlation is related to the mass of the cluster. We examine the relationship between $\Delta$Y and the absolute integrated magnitude and cluster concentration as indirect probes of the total mass of the cluster. The integrated magnitude presumably indicates that more stars (and hence more mass) are present in the cluster, while the concentration of the cluster is expected to be generally be higher in more massive clusters. The cluster concentration from \cite{Harris:2010} is a King-model central concentration, where the concentration = $\log$(r$_t$/r$_c$), where r$_t$ is the tidal radius and r$_c$ is the core radius. Core-collapsed clusters have a concentration value of 2.5 (\citealt{Harris:2010}).

We observe trends of helium abundance with the integrated magnitude of the cluster as well as the concentration of the cluster (both from \citealt{Harris:2010}). Both trends show positive correlations. For the spread of helium and absolute magnitude, we find an error-weighted Pearson correlation of --0.380 $\pm$ 0.040, with NGC 2298 as an outlier at an integrated V-band magnitude of only -6.0. For $\Delta$Y and cluster concentration, we determine a correlation of 0.232 $\pm$ 0.220; though the relation is scattered, the most highly concentrated clusters tend to be those with the larger spreads of helium. The evidence thus suggests cluster mass is a primary component in determining the amount of chemical enrichment in a cluster.

\begin{figure*} 
\includegraphics[width=0.95\textwidth]{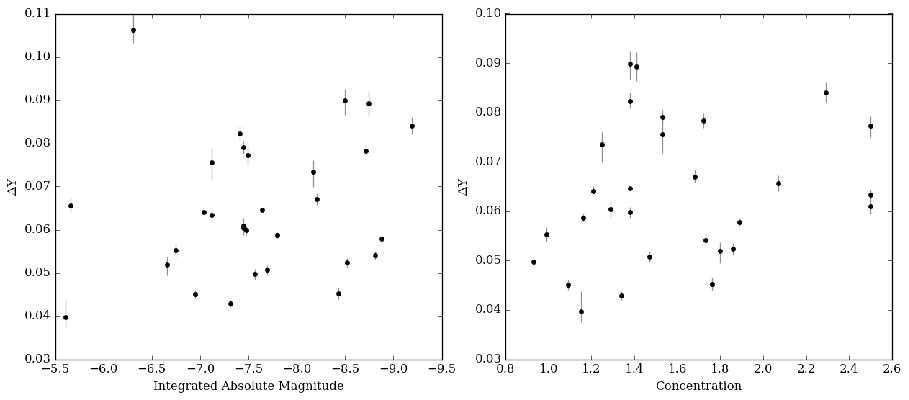}
\caption{The left panel shows the helium spread ($\Delta$Y) of the populations in the cluster plotted with the total integrated magnitude of the cluster (from \citealt{Harris:2010}). The right panel shows $\Delta$Y with respect to the velocity dispersion of the cluster (from \citealt{Harris:2010}).}
\label{Y_mass}
\end{figure*}

Figure \ref{Y_mass}, and previous work by \cite{Gratton:2010}, \cite{Milone:2014}, and \cite{Wagner-Kaiser:2016b}, also provide evidence that more massive clusters are able to achieve higher levels of helium enrichment. It may be that the more massive or more concentrated a cluster is, the more efficiently it is able to retain and reuse processed material. Under some formation scenarios, more massive clusters are presumed to better hold on to enriched ejecta from the evolving first population stars due to a deeper potential well, leading to generally higher helium enrichment of the forming second generation. In Section \ref{Forms}, we further investigate and discuss mass in context of formation scenarios of globular clusters.

\subsection{Population Proportion}\label{sec:prop}

\begin{figure*} 
\includegraphics[width=0.95\textwidth]{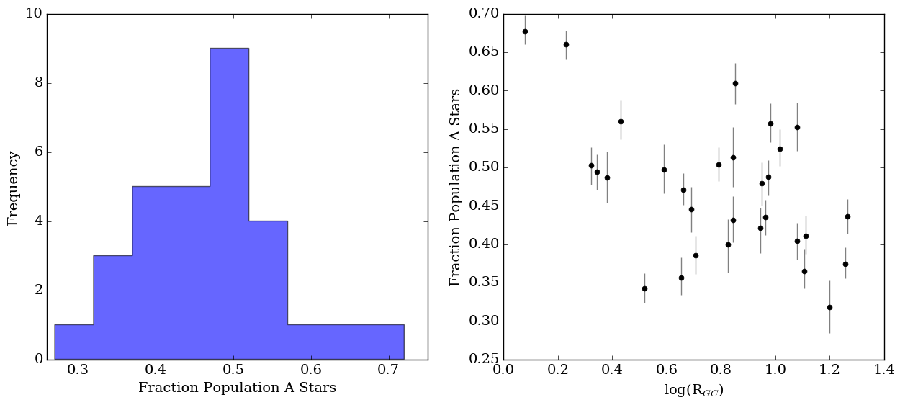}
\caption{Left: Distribution of the proportion for population A for the cluster sample. Right: The median of the posterior percentage of stars in the cluster likely belonging to population A, the first population of stars, is plotted with respect to the galactocentric distance of the cluster.}
\label{proportion}
\end{figure*}

We measure the proportion of stars in population A as the fraction of stars that likely belong to the stellar population with a lower helium fraction. We use photometry for 30 clusters and find a wide range of proportions from our Bayesian analysis, from $\sim$0.32 to 0.68 of the cluster belonging to population A, with a median of 0.47, as shown in Figure \ref{proportion}. 

In the right panel of Figure \ref{proportion}, we see the innermost clusters in the Galaxy tend to have a higher incidence of population A stars, while the outer clusters have lower fractions of population A stars. The relationship between the radial location of a cluster and its percentage of stars belonging to population A has a weighted correlation coefficient of --0.519$\pm$0.003. This trend is contrary to predictions from \cite{Bastian:2015} for a two-generation model, who suggest that the innermost clusters should have fewer population A stars under a two-generation formation scenario.

\begin{figure*} 
\includegraphics[width=0.95\textwidth]{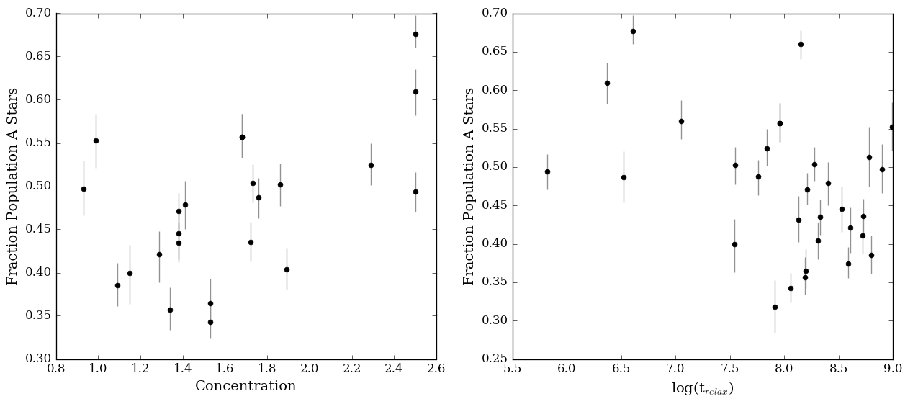}
\caption{Left: concentration of the clusters plotted with the percentage of stars belonging to population A. The weighted correlation coefficient is 0.552$\pm$0.006. Right: the relaxation times of the clusters compared to the percentage of stars belonging to population A, with a weighted correlation of --0.397$\pm$0.027.}
\label{proportion_RM}
\end{figure*}

We also examine the proportion of stars in population A with respect to the concentration and relaxation time of each cluster, as shown in Figure \ref{proportion_RM}. We find strong evidence that both the concentration and the relaxation time are related to the fraction of the cluster made up of population A stars, with Pearson error-weighted correlations of 0.552$\pm$0.006 and --0.397$\pm$0.027, respectively. As the concentration and the relaxation time are theoretically expected to be related to the mass of the cluster, this suggests that cluster mass also affects the amount of stars belonging to the chemically primordial or enriched populations in the cluster. Becayse the relaxation time also depends on the dynamics of the cluster, the correlation between relaxation time and proportion also leaves open the possibility that clusters are affected by their environment and orbital history.

We note the percentage of stars in a given population could be affected by the binary fraction of the cluster. Binaries were not included in this work because of the prohibitive computing time required to include binaries, but are expected to be largely classified as field stars in our statistical model. We do not find a trend between the binary fraction and the population proportion for the clusters, with a correlation of --0.035 $\pm$ 0.882 using the binary fractions determined by \cite{Milone:2012b}. Additionally, preliminary binarity tests of NGC 2298 (15\% binaries in core radius) and NGC 6656 (5\% binaries in core radius) in our sample find that the population proportion is unaffected by the inclusion of binaries, consistent within the 90\% Bayesian credible intervals.


\section{Discussion}\label{Discussion}

\subsection{Formation Models}\label{Forms}

We have presented the results of applying our Bayesian approach to a large sample of clusters using the methodology outlined and tested in \cite{Stenning:2016} and \cite{Wagner-Kaiser:2016}. From our results, based on our assumptions and the presently available models, we determine that clusters with higher helium enhancement tend to be more massive and more massive clusters appear to contain a higher proportion of first population stars. The range of helium mass fractions within each cluster tends to fall into the range of about 0.04 to 0.11.

We find evidence that the helium spread of a cluster is affected by mass, with more massive clusters reaching a greater difference in helium between two populations. The relationship between enrichment and mass has been predicted by several formation scenarios (AGB, FRMS, \citealt{Bastian:2015a, DErcole:2016}), and previous studies have found evidence of this relationship. We find additional evidence for this trend as well, incorporating measurements of 30 globular clusters, the largest consistently measured sample to date. 

More massive clusters may be better able to retain their first population of stars under a multi-generation scenario, which could explain the relationships we observe in Figure \ref{proportion_RM}. In any case, the unique orbital history would affect the extent of mass loss or stripping of each cluster, as inclination, radial location, and eccentricity all play a key role in cluster mass functions and dynamical evolution (\citealt{Webb:2014a,Webb:2015}). However, our result does not support previous studies that suggest under an AGB or FRMS scenario the inner clusters would have a smaller percentage of their first generation of stars (\citealt{Bastian:2015a}), as we see the opposite in Figure \ref{proportion}. It may be that mass is a stronger driver of the proportion of stars in each population than the radial location of the cluster.

The spread of proportions of the populations in the clusters we analyze in Section \ref{sec:prop} is contrary to the results suggested by \cite{Bastian:2015a}, who find a narrow range of enriched stars (i.e., population B) of 0.68$\pm$0.07 across 33 clusters based primarily on spectroscopy (with photometry for three of clusters, and both spectra and photometry for two clusters). We see a range of proportions from our Bayesian analysis, from 0.32 to 0.68 of the cluster belonging to population B, with a median of 0.47.

The difference in the ranges of proportions in the two studies likely arises due to the differences in star samples as well as methodology, discussed further in Section \ref{methods2}. With spectroscopy, the divide between populations is determined via locations of stars in the Na-O anti correlation, which is a continuous distribution for the current state of spectra for most clusters. Additionally, such a method may be affected by small number statistics. With photometry, the populations are discrete in the RGB (as in Figures \ref{UVCMDs1} and \ref{UVCMDs2}), and much larger samples of stars are available. However, biases may still be present and we expect future work to improve these issues.

In the future, combining spectroscopic and photometric data is the most promising route forward to determining more accurate analyses of population distributions. There is currently no consensus as to what parameter(s) affects the distribution of stars among populations in a cluster. Improved understanding on this subject will be highly informative on models of globular cluster formation.

\subsection{Methodologies}\label{methods2}

With renewed interest in globular clusters and new techniques for their analysis, a lively debate has arisen between research groups that prefer one method of analysis over another or one proposed formation mechanism to another.

The R-method for determination of helium has been around for some time (\citealt{Iben:1968}), relying on number ratios of HB and RGB stars in globular clusters, which would predict a higher helium content to lower the number of RGB stars. However, uncertainty in the calibration of this indirect method leads to large errors in the estimation of helium values, and studies often find any concrete conclusions thwarted by large errors (\citealt{Buzzoni:1983, Sandquist:2000, Zoccali:2000, Cassisi:2003}).

As knowledge of multiple populations has grown, so has the effort to examine more closely the chemical content of globular cluster stars. More recent studies have incorporated newer, more accurate high-resolution spectra or high-precision photometry to examine helium and other elements. Methods focusing on spectroscopy have slowly reduced measurement error and are beginning to concretely determine distinct chemical separations in a few clusters (e.g., \citealt{Carretta:2014}). Otherwise, in many spectroscopic studies, the exact location of the ``split" between populations (for example, in the Na-O anti-correlation diagram) has to been assumed in the continuous spread of data (\citealt{Carretta:2007, Carretta:2009a}). These studies are often restricted to the RGB or HB stars.

Another common approach of studying multiple populations is to examine a variety of CMD combinations to determine changes in various elements, including CNO and helium (\citealt{Piotto:2007,Bellini:2010,Milone:2012a,Piotto:2013,Milone:2013,Milone:2015,Nardiello:2015}, among others). This type of approach tests how well varying abundaces of helium and light elements (i.e. CNO) represent fiducial sequences of a cluster via incorporation of synthetic spectra, and exploring the fits at different evolutionary stages for various CMD combinations (\citealt{Milone:2009, Milone:2015}). However, these studies too must assume a reasonable location for the border between populations at a particular color in the globular cluster. This approach is able to incorporate many different aspects of the multiple population scenario, but generally focus on individual clusters with unique assumptions rather than the GC population as a whole.

Our method explores a large swath of parameter space to determine the most likely description of multidimensional photometric data; it is statistically driven and robustly objective. Like other studies, our underlying assumptions include the accuracy of the models, typical initial mass functions, existence of field stars, etc. However, we do not assume values for helium or the delineation of populations in the CMD, allowing the statistical framework to determine these instead. We don't look at one particular CMD set, but simultaneously consider all possible CMDs at once in a multidimensional fit. We do so for a large number of clusters to obtain a preliminary picture of the possible driving forces of multiple population characteristics. Future development will also allow us to directly incorporate spectroscopic information.

However, the primary drawback of our approach is that the [$\alpha$/Fe] ratio, which can mimic changes in helium in the morphology of the CMD, is not currently a parameter in the statistical framework. Variations in C, N, and O may also affect differences in the specific HST filters used herein and the currently available models are not able to account for these variations. Although these variations are thought to be correlated with helium (\citealt{Mucciarelli:2014}), because we do not take into account variations in light element or [$\alpha$/Fe] abundance variations among populations within a cluster, we must view our results as upper limits on changes in helium. We are presently limited to estimating the upper limit of $\Delta$Y, which may incorporate a spread in [$\alpha$/Fe] in conjunction with a spread in Y. 

In the context of our results, this means that for some clusters, the estimated $\Delta$Y may be primarily driven by differences in light element abundances. In other cases, variations of other elements may only be a minor effect. Until a sufficient suite of isochrones varying in C, N, and O are available in the appropriate UVIS filters, we will not be able to disentangle the two. We look forward to incorporating these in our analysis once possible. 

Additionally, we stress the caveat on these results by reiterating that our sample is made up of only clusters that appear to harbor two dominant populations. Further work is needed to determine whether or not clusters with more stellar populations follow the same trends.

As the exploration of various techniques and solutions to the multiple populations advances, we will continue to expand and improve our approach, continuing to compare to alternate techniques.

\subsection{Model Inconsistencies}\label{modelincon}

\begin{figure*}
\includegraphics[width=\textwidth]{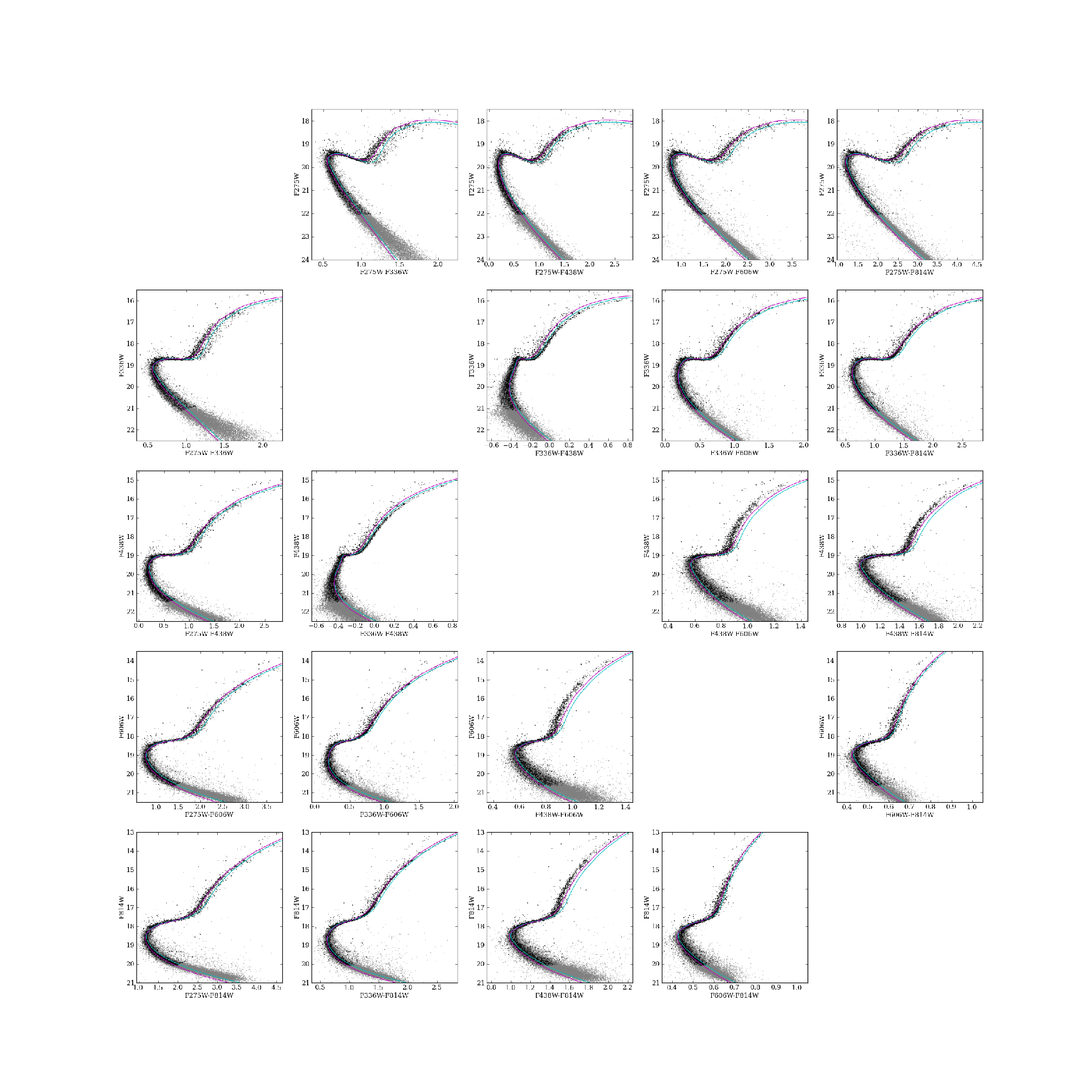}
\caption{A grid of all possible CMDs of NGC 5272 from the five filter UVIS and ACS treasury data (F275W, F336W, F438W, F606W and F814W), wavelength increases moving from left to right and top to bottom. All stars are shown in black and the subsample of stars fit with BASE-9 is shown in gray. The BASE-9 determined model fits are shown as isochrones constructed from median values of the MCMC sampling, with population A shown in cyan and population B in magenta.}
\label{NGC5272}
\end{figure*}

In our Bayesian method, because we assume the input models fully represent cluster behavior, our results will only be as descriptive as these models. As the theoretical models improve (or include estimates of theoretical uncertainty), so will the accuracy at which we characterize the clusters. However, we can use our methodology to provide valuable feedback on the models by comparing where the behavior of the CMD deviates from the predicted behavior from the isochrones. We use NGC 5272 as a sample case, with a grid of CMDs shown in Figure \ref{NGC5272}.

For most clusters, in comparing the observed photometry to the isochrones, we note inconsistencies in the theoretical models in the ultraviolet wavelengths, and disagreements between the ultraviolet and visual wavelengths. Here, we summarize the main inconsistencies that we see when examining many clusters. While fitting any of these CMDs individually may lead to a more aesthetically pleasing fit, the end goal is to have models fit the observed data in all filters \emph{simultaneously}. We hope that by fitting the observed photometry in all filters, we can help pinpoint where in the CMD that the theory needs to be improved.

In the F275W -- F336W CMD, the theoretical models predict a much more linear lower MS than we see in the data. The curvature of the observed sequence stretches more redward than the models at faint magnitudes. Additionally, metal-rich clusters have sub-giant branches (SGB) that are ``too short", in that the difference in color between the MSTOP and the bottom of the RGB is too small. For more metal-poor clusters, this appears to be less of an issue.

The F336W -- F438W CMD also sometimes has issues with lower MS curvature, much the same as the F275W -- F336W CMD, where the models predict a linear lower MS that the data do not follow. Otherwise, this combination of filters tends to match the observed data fairly well.

The more metal-poor clusters appear to be fit best in the F438W -- F606W CMD, but the majority of clusters tend to have the observed stars bluer than predicted by the isochrone. This is especially true along the RGB, and the models again predict the lower MS to be linear; however, this is not observed in the photometry.

As may be expected, the F606W -- F814W CMDs fit the data better than other filters for most clusters. However, in some cases the models predict that the base of the RGB is redder than the observed data. Sometimes this persists up the length of the RGB as well. 

Additionally, we find that the most metal-rich isochrones ([Fe/H]$\lesssim$--1.0) tend to be particularly troublesome in the RGB at ultraviolet wavelengths due to uncertainties in atmospheric and opacity models.

The ``push and pull" between different filters in the models can affect the resulting isochrone parameters when fitting the photometry. We see this in comparing our results in this work using five filter photometry to that of \cite{Wagner-Kaiser:2016b}, who use only two filters (F606W and F814W). Between the two studies, we observe systematic biases in distance and age. Essentially, the statistically optimal isochrone based on visual wavelengths is not equivalent to the most appropriate fit for the ultraviolet wavelengths, due to model mismatch, as has been investigated for a single population by \cite{Hills:2015}. We expect that, in the future, when we are able to incorporate corollary information from abundance measurements, primarily CNO, that many of the issues and disagreements will improve. Following this, we hope that results from using different combinations of filters will lead to more consistent results.


\section{Conclusions}\label{Conclusions}

We analyze 30 Galactic globular clusters that appear to harbor two stellar populations. Our analysis simultaneously obtained an age, distance, and extinction for each cluster. At the same time, a helium value is determined for each of the two populations within the cluster as well as the proportion of stars that belong to each stellar population. From this analysis, we draw the following conclusions:

1. We present the largest sample of 30 uniformly analyzed Galactic globular clusters under the multiple population scenario, using a Bayesian statistical technique.

2. The helium enrichment differences between the two populations of these clusters tends to fall into a narrow range of $\Delta$Y $\sim$ 0.04 to 0.11. In all cases, we expect that our estimates of the helium differences are upper limits.

3. Similar to previous work, we see a strong indication that more massive clusters have a higher spread of helium enrichment than less massive clusters, possibly due to a deeper gravitational potential. Under an AGB or FRMS formation scenario, a deep gravitational potential well would assist the cluster in holding on to enriched material. 

4. Correlations are observed between of the fraction of stars in population A with both mass and radial location in the Galaxy. This suggests that environment and orbital history may play a significant role in the multi-faceted picture of factors driving multiple population characteristics in clusters.

 
\section*{Acknowledgments}

We are grateful to the referee, whose thoughtful comments greatly improved the paper. Based on observations with the NASA/ESA Hubble Space Telescope obtained at the Space Telescope Science Institute, which is operated by the Associations of Universities for Research in Astronomy, Incorporated, under NASA contract NAS5-26555. These observations are associated with program GO-13297.23-A. This material is based upon work supported by the National Aeronautics and Space Administration under Grant NNX11AF34G issued through the Office of Space Science, and through the University of Central Florida's NASA Florida Space Grant Consortium. DS work was supported by NSF grants DMS 1208791. DvD was partially supported by a Wolfson Research Merit Award (WM110023) provided by the British Royal Society and by Marie-Curie Career Integration (FP7-PEOPLE-2012-CIG-321865) and Marie-Skodowska-Curie RISE (H2020-MSCA-RISE-2015-691164). Grants both provided by the European Commission.


\bibliographystyle{mn2e}
\bibliography{TpaperII}
\clearpage

\label{lastpage}

\end{document}